\documentclass[reprint,aps,pra,amsmath,amssymb,showpacs]{revtex4-1}
\usepackage{graphicx}
\usepackage{dcolumn}
\usepackage{bm}
\usepackage{color}
\usepackage{upgreek}
\usepackage{latexsym}

\begin{document}

\title{Probing dephasing time of crystals via spectral properties of  high-harmonic generation }

\author{Tao-Yuan Du }\email{duty@cug.edu.cn} \affiliation{School of Mathematics and Physics, China University of Geosciences, Wuhan 430074, China}

\date{\today}

\begin{abstract}
Coherent time is a characteristic time in the extreme nonlinear optics regime and thus generally introduced as the dephasing time in the simulations of the solid-state high-harmonic generation. This characteristic time linked with the coherent decay of quantum trajectory controls the emergence of the spectral splittings in the high harmonic spectroscopy, which can been attributed to the temporal interference between two adjacent harmonic emission channels. To reproduce the harmonic peaks and spectral splittings, a temporal two-slit interference model in the momentum space is introduced.  In addition, mechanisms of spectral splitting provide a state-of-the-art avenue to probe the characteristic elapse between electron and hole. Based on the subfemtosecond resolution of the spectral interference opening and closing zones in the wavelength-dependent high-order harmonic spectra, we also propose an alternative scheme to probe the dephasing time of crystals by the feasible laser pulses in present experimental setups.
 
\end{abstract}

\maketitle

\section{Introduction}
Since its first observation in rare gases three decades ago, high-order harmonic generation (HHG) has become the foundation for attosecond physics through a series of advances \cite{Ferray}. Attosecond pulses generated by gas-phase HHG provide a powerful tool for the probe of ultrafast dynamics in atoms, molecules, and biological systems \cite{Peng,Hentschel,Smirnov,Worner}. Given the potential as a bright, compact, controllable source of extreme ultraviolet radiation and the promise of applying HHG spectroscopic techniques to correlated quasiparticle dynamics in condensed matter \cite{Kruchinin, Pronin}, the realization of efficient HHG from transparent crystals has inspired enormous interests \cite{Ghimire,Vampa,Ndabashimiye,You_NC,Liu_hanzhe,Yoshikawa,You_NP,Luu} and been demonstrated as an alternative way to study properties of the bulk materials, such as the reconstruction of the energy bands and the Berry curvature \cite{Vampa_recon,Banks_Berry,Luu_Berry,Yangfang1,Yangfang2,Yangfang3}. 

The physical picture of HHG in solids is well described by a three-step model in reciprocal space: (i) an electron wave packet tunnels to conduction band; (ii) without scatterings, the electron driven by the electric field $\mathbf{F}$ is expected to map the dispersion of conduction band at a constant rate in momentum space, $ d\mathbf{k}/dt = \mathbf{F}$; (iii) then the electron recombines back to the hole remained in the original valence band, leading to the burst of high harmonics with energy equal to the energy gap between valence and conduction band states at the crystal momentum of recombination \cite{Du1,Du3,Hawkins,Ikemachi,Du2}. One could find that two families of the electron trajectories with different excursion times contribute to two channels emitting the same harmonics. These trajectories, or quantum paths, are distinguished as short and long, reflecting the duration of the electron's excursion time in conduction band, and forming the dynamical Bloch oscillations of the electron wave packet. Thus, the quantum paths that electrons are driven toward the boundary of  Brillouin zone (BZ) and pulled back the center of BZ are called as \textit{short} and \textit{long} trajectories respectively. 

However, experimental observation of these oscillations in conventional bulk solids is difficult due to the strong electron-particle scatterings which lead to the short intraband population relaxation times compared with Bloch oscillation cycle \cite{Rossi1,Rossi2,Kruchinin}. The electron-particle interaction with an environment will destroy the interband coherence which can be phenomenologically characterized by the dephasing time, reflecting the phase relaxation times at a few femtoseconds timescale, which is comparable to the excursion time of the electron-hole pair driven by the developed mid-infrared (MIR) laser pulses and can be evaluated from the electron-particle interaction integrals \cite{Landau,Kaganov}. 
In addition, the dephasing time dramatically changes the high-harmonic spectroscopy in theoretical simulations and results in the spectral splittings, which had been observed experimentally without any insightful discussions \cite{Ndabashimiye,You_NC}. The spectroscopic fine structure of HHG are central to extract the dynamics of the electron-hole pair in condensed matter.

In this work, we find that the spectroscopic characteristics delicately reflect the ratio of the dephasing time and the excursion time of electron-hole pair, which is used to probe the dephasing time of crystals. Theoretically, we establish the subfemtosecond resolution of the dephasing time with the measurable spectral splitting in harmonic spectroscopy.

\section{THEORETICAL METHODS}
In our simulation, the interaction of the laser light with the wurtzite ZnO (\textit{w}-ZnO) crystal is modeled by solving the multiband density matrix equations (DMEs), which can be also found in Refs. \cite{Vampa1,Vampa4,Du5,Vampa5,zhangxiao}. The multiband DMEs are written as
\begin{equation}\label{E1}
	\dot{n}_{m} = i \sum_{m' \neq m} \Omega_{mm'} \pi_{mm'}e^{iS_{mm'}} + c.c.,
\end{equation}
\begin{equation}\label{E2}
	\begin{split}
		&  \dot{\pi}_{mm'} =  - \frac{\pi_{mm'}}{T_{2}} + i\Omega_{mm'}^{*}(n_{m} - n_{m'})e^{-iS_{mm'}} \\
		& + i \sum_{m'' \notin \{m,m'\}}(\Omega_{m'm''} \pi_{mm''}e^{iS_{m'm''}} - \Omega_{mm''}^{*} \pi_{m'm''}^{*} e^{-iS_{mm''}}).
	\end{split}
\end{equation} 
Here, $n_{m}$ is the population for band $m$ and is subject to the constraint $\sum _{m} n_{m} = 1$. For simplicity, we have dropped the input $(\mathbf{k}, t)$ in Eqs. (\ref{E1}) and (\ref{E2}).  Further,  $(\mathbf{k}, t) \equiv \mathbf{k} +  \int_{-\infty}^{t}\mathbf{F}(t')dt'$ is the semiclassical equation \cite{Huangkun,Ashcroft}. $S(\mathbf{k}, t) = \int_{-\infty}^{t}\varepsilon_{g}(\mathbf{k}, t')dt'$ is the classical action, and $\varepsilon_{g}$ is the band gap between the coupled energy bands.  $\Omega (\mathbf{k},t) =  \mathbf{F}(t)\mathbf{d}(\mathbf{k}, t)$ is the Rabi frequency and $\pi(\mathbf{k}, t) $ is related to interband polarization term $ \textbf{p}(\mathbf{k}, t) = \mathbf{d}(\mathbf{k}, t)\pi(\mathbf{k}, t)e^{iS(\mathbf{k}, t)} + c.c.$ $T_{2}$ is the dephasing time we are concerned with and introduced to account for coupling to a phonon bath and impurities and for electron-electron scattering.
The nonlinear currents induced by the light-solids interaction can be calculated as
\begin{equation}\label{E3}
\textbf{j}_{ra}(t) = \sum_{m=c, v} \int_{BZ} v_{m}(\mathbf{k}, t)n_{m}(\mathbf{k},t)d\mathbf{k},
\end{equation}
\begin{equation}\label{E4}
\textbf{j}_{er}(t) = \frac{d}{dt} \int_{BZ} \textbf{p}(\mathbf{k}, t)d\mathbf{k},
\end{equation}
where the band velocity $v_{m}$ is defined by $  \nabla_{\mathbf{k}}E_{m}(\mathbf{k})$. The high-harmonic spectrum is obtained from the Fourier transform (FT) of $\textbf{j}_{total} = \textbf{j}_{ra} + \textbf{j}_{er}$, as $|FT\{\textbf{j}_{total}\}|^{2}$. We obtained \textbf{d}(\emph{k}) and the band energies $\emph{E}_{m}(k)$ by using the Vienna ab initio simulation package (VASP) code. The details of the energy bands and transition dipole elements can be found in the Appendix A. The linearly polarized laser field is oriented along the $\Gamma M$ direction. To exclude the complexity of  more than two emission channels within a half cycle induced by the dynamical Bloch oscillation across the edge of BZ, the laser intensity adopted here is moderate \cite{Hawkins,Du3}.

To obtain the excursion time of electron-hole pair in solid crystals driven by the laser fields, the electron-hole recollision model is used. The saddle-point analysis is an efficient approach to extract the excursion times of electron and hole in their respective energy bands. Considering the laser parameters are used here, the interband contribution dominates the generation of harmonics in the primary plateau zone \cite{Vampa1,Du_inhomo}. Hence, hereafter we focus on the HHG contributed by interband current. For simplicity, a two-band integral form which is equivalent to Eq. (\ref{E2}) without any approximation can be rewritten as
\begin{equation}\label{E5}
  \pi(\mathbf{k}, t)  = - i \int_{-\infty}^{t}\Omega(\mathbf{k}, t') \Delta n(\mathbf{k}, t) \zeta(T_{2}, t-t')  e^{-iS(\mathbf{k}, t)}dt',
\end{equation}
\begin{equation}\label{E6}
\zeta(T_{2}, t-t') = e^{-(t-t')/T_{2}},
\end{equation}
where $\zeta(T_{2}, t-t')$ shows the coherence decay caused by the scatterings \cite{Lewenstein} and $\Delta n $ is the population difference between electron and hole. By substituting Eq. (\ref{E5}) into Eq. (\ref{E4}) and then performing FT, the effect of coherent decay be embodied in the interband emissions as follow: 
\begin{equation}\label{E7}
\begin{split}
\textbf{S}_{er}(\omega) = &  -\omega^{2} \int_{BZ} d\mathbf{k}\int_{-\infty}^{\infty}dt e^{-i\omega t} \lbrack  \int_{-\infty}^{t} \mathbf{d^{*}}(\mathbf{k}, t')  \\
                                           &\ \times \Omega(\mathbf{k},t') \Delta n(\mathbf{k}, t')\zeta(T_{2}, t-t') e^{i\phi(\mathbf{k}, t, t')} + c.c. \rbrack, 
\end{split}
\end{equation}
where the classical action phase is represented by 
\begin{equation}\label{E8}
\phi(\mathbf{k}, t, t') = - \int_{t'}^{t} \epsilon_{g}(\mathbf{k}, t'')dt'' - \omega t.
\end{equation}

The saddle-point equations obtained by the first derivative of $\phi$ can be described as 
\begin{equation}\label{E9}
 \int_{t'}^{t} v_{g}(\mathbf{k}, t'')dt'' = 0,
\end{equation}
\begin{equation}\label{E10}
\frac{d\phi}{dt'}  = \epsilon_{g}(\mathbf{k}, t') = 0,
\end{equation}
\begin{equation}\label{E11}
\frac{d\phi}{dt}  =  \epsilon_{g}(\mathbf{k}, t) = \omega,
\end{equation}
where $v_{g} = \nabla_{\mathbf{k}}\epsilon_{g}(\mathbf{k})$ is the group velocity difference between electron and hole. To solving the saddle-point equations, we consider the electron trajectories initially located around the top state of valence band (including 5\% area of the BZ) because the electron ionization mainly occur in the part of the BZ where the corresponding transition dipole is maximal \cite{Du5}. Note that the choice for the larger area of valence band states in the BZ will result in similar conclusions.

\begin{figure*}[htbp]
	\centering
	\includegraphics[width=14.5 cm,height=10 cm]{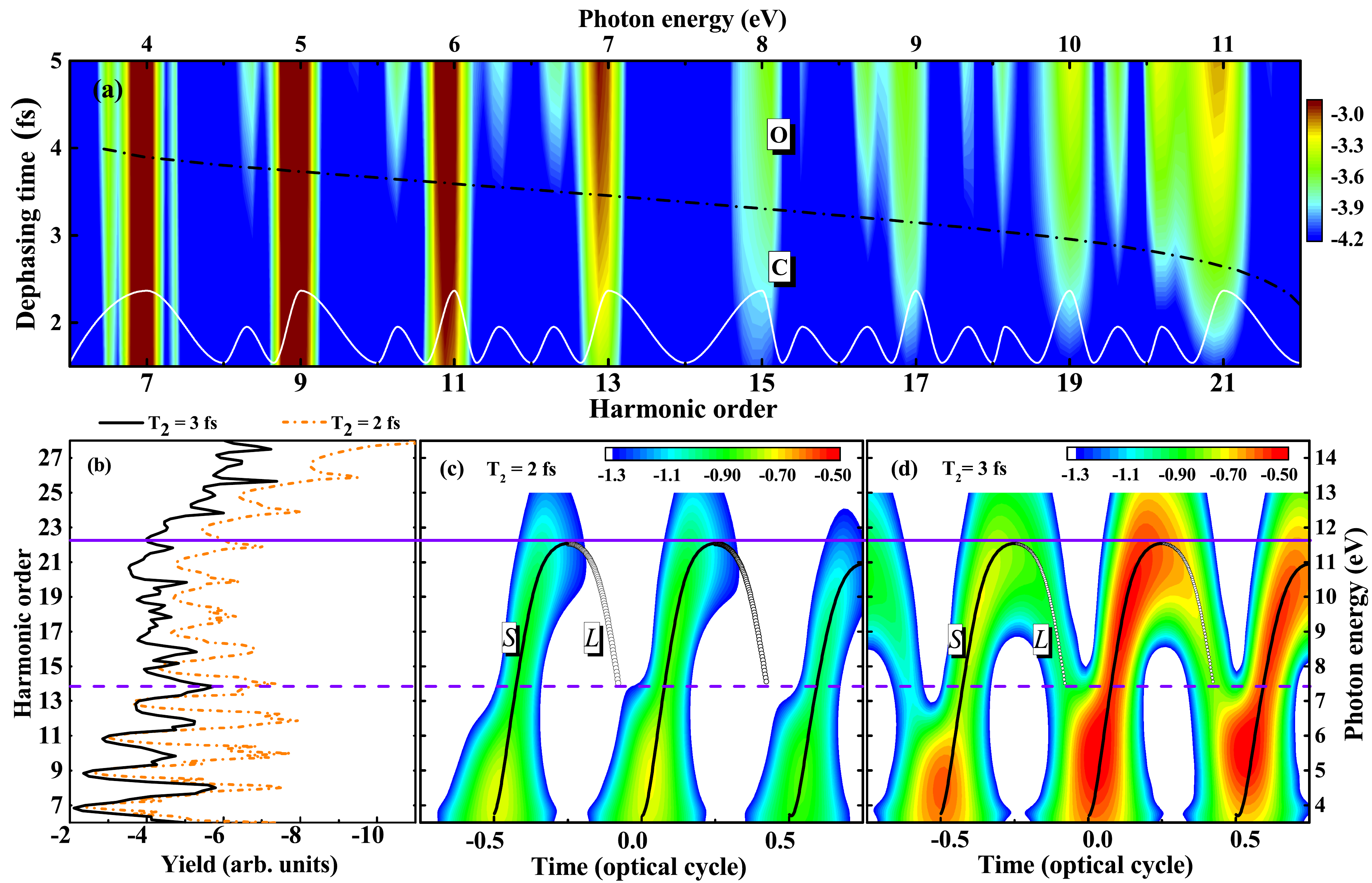}
	\caption{(Color online) (a) High-harmonic spectra as a function of the dephasing time with constant laser intensity 2.84 TW cm$^{-2}$. The laser pulses with a sin$^{2}$ envelope, duration eight optical cycles and  laser wavelength 2400 nm, corresponding to a laser period 8 fs. (b) High harmonic spectra with dephasing times 3 fs (black solid curve) and 2 fs (orange dash-dotted curve). (c) and (d) are the time-frequency analysis of the HHG spectra in (b) with dephasing times 2 fs and 3 fs respectively. The predicted boundary between the interference opening (\textbf{O}) closing (\textbf{C}) zones is shown by the black dash-dotted line in (a) and violet dash line in (b).  The white curve in (a) shows the positions of the harmonic peaks and spectral splittings which are predicted by the temporal two-slit interference contributed by the short and long trajectories. The semiclassical analysis shown by black line and circle in (c) and (d) is separated into the short (\textit{S}) and long (\textit{L}) trajectories according to the excursion time of electron. The color scale is logarithmic.}
	\label{Fig1}
\end{figure*}

The classical electron trajectory starting at ionized moment $t'$ and ending at recombined moment $t$ satisfies Eq. (\ref{E9}) . Note that the electron and hole wave functions are Bloch waves and can be delocalized in solids. Considering the substantial overlap between the subwaves of the electron-hole pair, we will relax the recollision condition of zero displacement. The electron trajectories with different ionization moments are coherent with each other and the ionization event dominantly occur around the peak of the electric field. An electron ionized at moment $t'$ usually recombines with hole at several moments $t$. For the effect of the decoherence term $\zeta$ in Eq. (\ref{E7}), high-order recollision trajectories with the excursion time $\tau = t - t'$ larger than a half optical cycle of pulse will be suppressed dramatically. Consequently, only the first electron trajectory characterized by the excursion time lower than a half cycle is considered in our classical analysis. And the electron excursion time $\tau_{q} $ for the high harmonic $q\omega_{0}$ ($\omega_{0}$ is the laser frequency, and $q$ is the order) increases with the growing order. 

\section{RESULTS AND DISCUSSIONS}
Firstly, we investigate the role of the coherent time in the HHG from solids. Figure \ref{Fig1}(a) shows the high-harmonic spectra as a function of the dephasing time and two representative HHG spectra are presented in Fig. \ref{Fig1}(b). One can observe two features: (i) the yield for a certain harmonic is enhanced monotonically with increasing dephasing time and this enhanced effect of yield becomes more significant for the higher harmonics; (ii) spectral splitting and redshift.  

\begin{figure}[htbp]
	\centering
	\includegraphics[width=8 cm,height=5 cm]{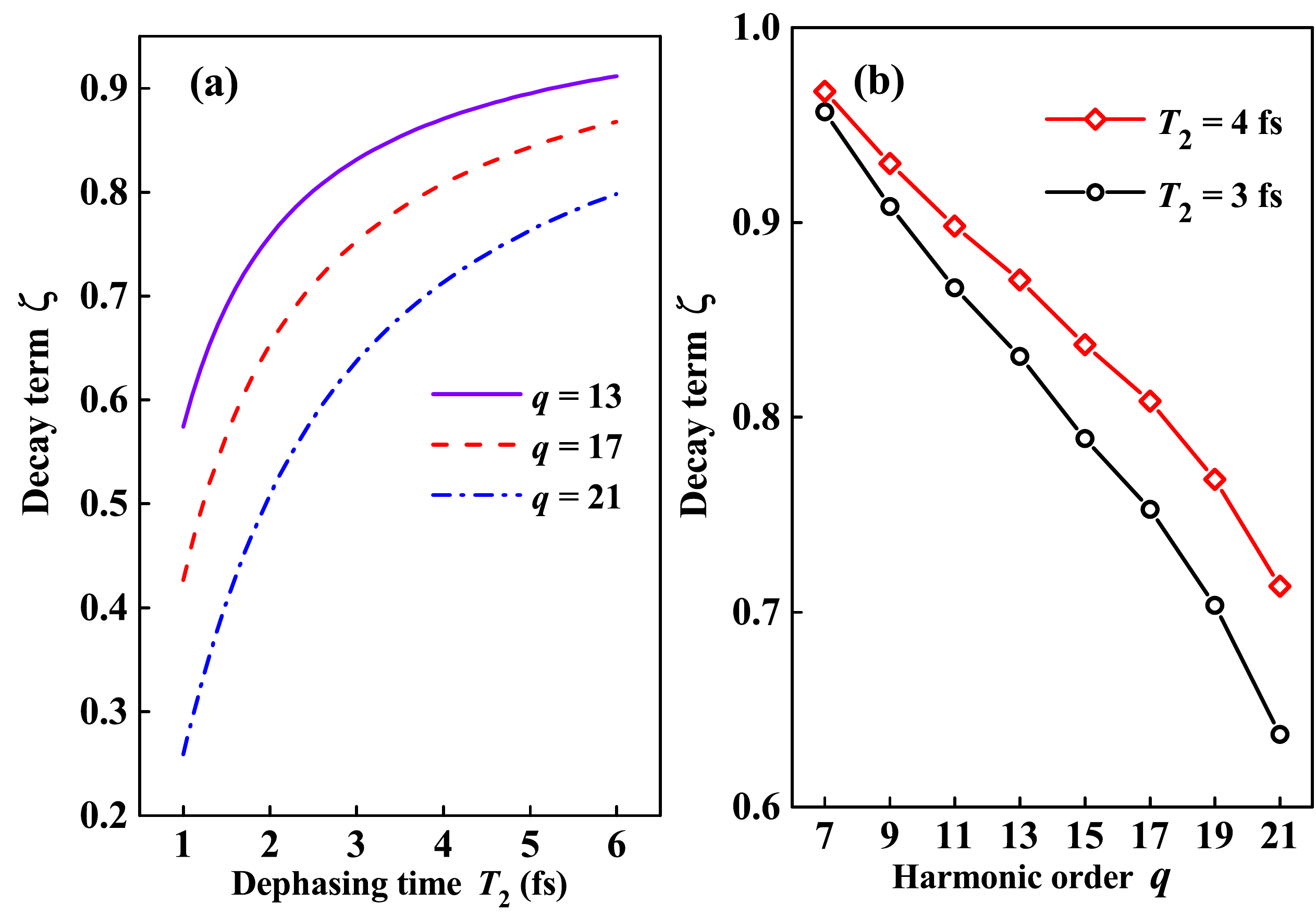}
	\caption{(Color online) (a) The decay term $\zeta$ in Eq. (\ref{E6}) with growing the dephasing time. (b) The decay term $\zeta$ as a function of harmonic order $q$.          }\label{Fig2}
\end{figure}

As shown in Fig. \ref{Fig1}(b), one can clearly discover the first feature by taking a contrast between two representative HHG spectra in which the dephasing times $ T_{2} =$ 2 and 3 fs are adopted respectively. To make an insight into the first feature, the clear physical comprehension and meaning can be captured by assessing the decoherence term $\zeta$ (less than one) in Eq. (\ref{E6}). As presented in Fig. \ref{Fig2}(a), the term $\zeta$ with a ratio of the constant excursion time $\tau_{q}$ and the growing $T_{2}$ leads to the slow decay (larger $\zeta$ value) for the interband polarization in Eq. (\ref{E7}), which implies that a smaller friction force acts on the electron or a larger elapse between two scattering events and then results in the enhanced harmonic yields. Similarly, considering the excursion times $\tau_{q}$ with growing $q$ and the $T_{2}$ are simultaneously increasing, the values of the decay term $\zeta$ for the higher harmonics will increase more rapidly with growing dephasing time, as shown in Figs. \ref{Fig2}(a) and \ref{Fig2}(b). Consequently, compared with the lower harmonics, the decay rates of the higher harmonics are slower when the dephasing time is growing. The slower decay rates (larger $\zeta$ values) in Eq. (\ref{E7}) give rise to the greater yield enhancement for the higher harmonics. In Figs. \ref{Fig1}(a) and  \ref{Fig1}(b), this feature can be intuitively observed in the range from 10 to 11 eV compared with the range from 8 to 9 eV. The dephasing time in laser-driven solid materials could be controlled by the means of band-gap engineering. Thus, the mechanism of harmonic yield modulated by dephasing time will provide new route to control the generation of high harmonic in solids. 

The second feature (spectral splitting and slight redshift of harmonic peaks with varying dephasing time) could also be found in Fig. \ref{Fig1}(a) and the black solid curve of Fig. \ref{Fig1}(b). In the spectroscopic range enclosed by two paralleled violet lines, both spectral splitting and redshift are clearly seen. However, the spectroscopic range with lower energy characterizes negligible spectral splitting and redshift. Note that effect of the complex channel interference caused by multiband and higher laser intensity on the second feature will be discussed in the Appendix B. Therefore, the HHG plateau can be divided into two parts according to the emergence and absence of the second future. To understand the second feature, we will turn to the view of \textit{short} and \textit{long} trajectories. Keeping the above-mentioned first trajectory in mind, we resolve two re-colliding times, less or more than a quarter optical cycle, corresponding to \textit{short} and \textit{long} trajectories for $q$-th harmonic and denoted as $\tau_{q}^{\textit{S}}$ and $\tau_{q}^{\textit{L}}$, respectively. 

For the secondary feature, we first reveal the mechanisms for the emergence of the spectral splitting caused by the interference between two emission channels or slits. In the reciprocal space, the ionized electrons are accelerated and will acquire a shift of the crystal momentum during the excursion time. The acquired momenta of the electron for the \textit{short} and \textit{long} trajectories are determined by their duration of the excursion times, and they are indicated as $\Delta k^{S}_{q}$ and $\Delta k^{L}_{q}$, respectively.  However, the phase-breaking events described by the characteristic time $T_{2}$ could destroy the process of the momentum accumulation. Within this elapse, the time-dependent crystal momenta we are concerned with can be referred as 
\begin{equation}\label{E12}
(\mathbf{k}^{S(L)}, t) \equiv \mathbf{ k_{0}} + \Delta k^{S(L)}_{q} = \mathbf{ k_{0}}  + \int_{t'}^{t'+\tau_{q}^{\textit{S(L)}}}\mathbf{F}(t)dt,
\end{equation}

\begin{equation}\label{E13}
(\mathbf{K}, t) \equiv \mathbf{k_{0}}  +  \int_{t'}^{t'+T_{2}}\mathbf{F}(t)dt,
\end{equation}
where the time-dependent \textit{short} and \textit{long} trajectories in the reciprocal space will be obtained by Eq. (\ref{E12}) and presented by the white line and green circles in Figs. \ref{Fig1}(c) and \ref{Fig1}(d) respectively.  $(\mathbf{K}, t)$ shows the whole semiclassical trajectory during the scattering-free elapse $T_{2}$.

\begin{figure}[htbp]
	\centering
	\includegraphics[width=6 cm,height=4.5 cm]{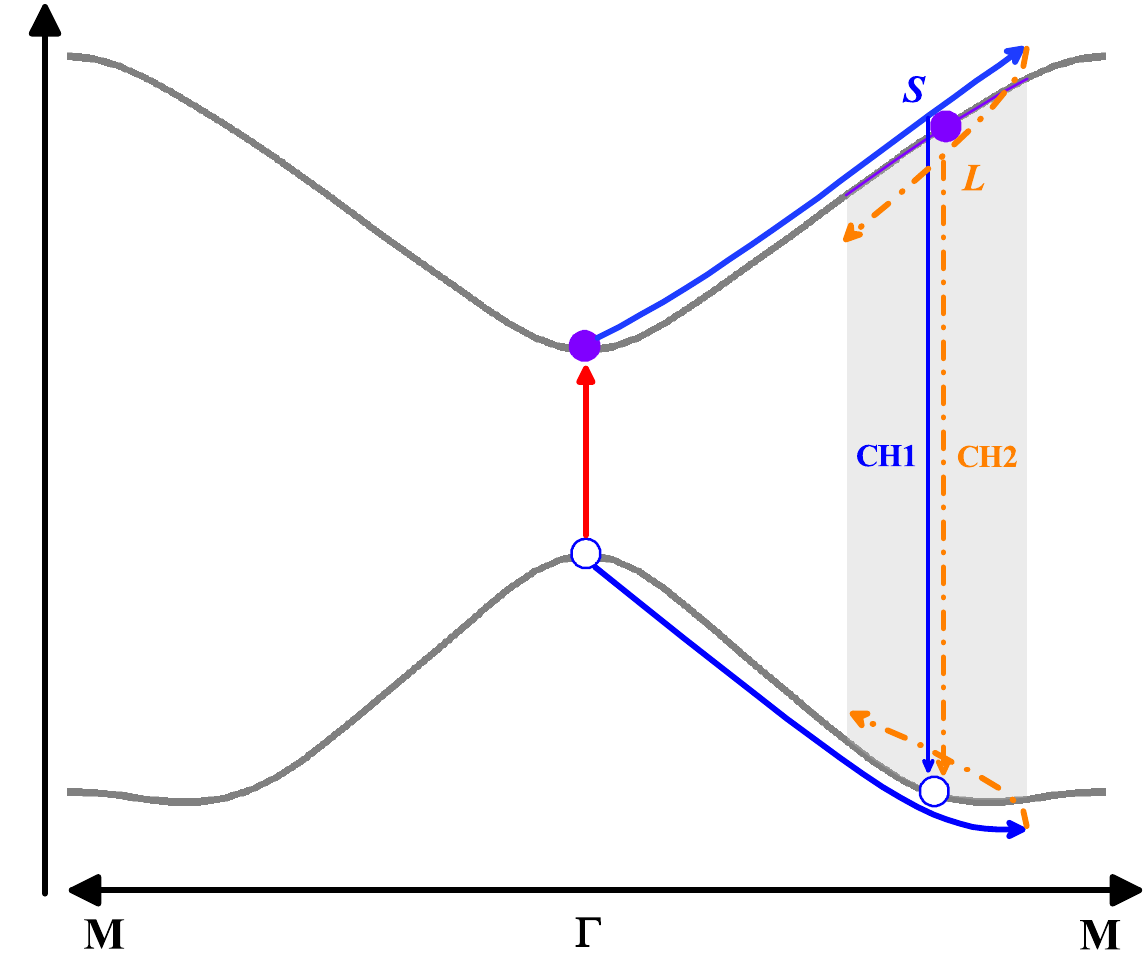}
	\caption{ (Color online) Schematic of two-slit interference contributed by the \textit{short} (\textit{S}) and \textit{long} (\textit{L}) trajectories. An electron around $\Gamma$ point of valence band is excited into a conduction, propagated within that band and then recombined with its parent hole in the original valence band. Ignoring the scattering effect, the electron/hole in its band shows the periodic Bloch oscillations and would be pulled back to the $\Gamma$ point within a half optical cycle. The electron dynamics toward the boundary is denoted as \textit{short} trajectory (blue solid curve) and gives rise to the emission channel CH1 for certain harmonic (the first slit), while the return electron dynamics contributes to \textit{long} trajectory and the additional emission channel CH2 for the same harmonic (the second slit). An appropriate dephasing time deduced by the scattering effect would support the existence of a part of \textit{long} trajectories, as shown by the orange dash-dotted curve. The shadow shows the opening zone of the two-slit interference between CH1 and CH2, and a counterpart zone at the opposite side of the BZ is equivalent (not shown).}\label{Fig3}
\end{figure}

To further resolve the quantum trajectories from the spectroscopy of HHG, we also show the time-frequency analysis of the two representative HHG spectra in Figs. \ref{Fig1}(c) and \ref{Fig1}(d) \cite{Chandre}. One can observe that the semiclassical paths reach a good agreement with the quantum trajectories \cite{Du1}. For the excursion time $\tau_{q}^{\textit{S}}$ shorter than $\tau_{q}^{\textit{L}}$, the role of decay term $\zeta$ in Eq. (\ref{E7}) will mainly dampen the \textit{long} trajectory, as shown by the time-frequency analysis in Fig. \ref{Fig1}(c) compared with that in Fig. \ref{Fig1}(d). When the dephasing time $T_{2}$ is less than a quarter of optical cycle, the \textit{long} trajectory will be faded away and one emission channel or slit is turned off, which results in the absence of the spectral splitting. Otherwise, the \textit{long} trajectory shall survive under the condition that $T_{2}$ greater than a quarter of optical cycle. Thus, the two-slit interference contributed by two emission channels is turned on. As shown by the black dash-dotted curve in Fig. \ref{Fig1}(a) and the violet solid line in Fig. \ref{Fig1}(b), the boundary between the spectral interference opening and closing zones is calculated by  
\begin{equation}\label{E14}
\eta = \epsilon_{g}(\mathbf{K}, t),
\end{equation}
where $\mathbf{K}$ is the accumulated crystal momentum in Eq. (\ref{E13}) during the elapse of $T_{2}$. In a word, the ratio of  $\tau_{q}^{\textit{S(L)}}$ and $T_{2}$ would turn on or off the one of two slits and then control the spectral interference range according to the Eqs. (\ref{E12}) and (\ref{E13}), as presented by the shadow zone in Fig. (\ref{Fig3}).

Within a half optical cycle, the understanding of the spectral interference can be formalized by considering the time-domain interference between two adjacent harmonic bursts (two slits) contributed by \textit{short} and \textit{long} trajectories respectively, as shown by the Fig. (\ref{Fig3}). The harmonic bursts generated by \textit{short} and \textit{long} trajectories have different amplitudes and phases, which can lead to constructive or destructive interference. The dynamic phase difference between two adjacent bursts for certain harmonic: $\Theta = \int_{t_{1}}^{t_{2}} \epsilon_{g}(\mathbf{k},t)dt$, where $t_{1}$ and $t_{2}$ are the moments when a certain harmonic component is generated by the CH1 and CH2 channels, respectively. In consequence, one harmonic component from the channel CH1 constructively interferes with its counterpart from the channel CH2, causing spectral interference of spectroscopy and split in the pattern of the HHG spectra varying with dephasing time \cite{You_NC}. The peaks of spectral splitting and $q$-th harmonic can be predicted by  
\begin{equation}\label{E15}
q =\frac{ (2l-1) - \Theta/\pi} {2\Delta t/T_{0} },
\end{equation}
$\Delta t = t_{2}-t_{1}$ is the temporal slit gap. $l$ is a positive integer and $T_{0}$ is the optical cycle. If the \textit{long} trajectory is absence, only the odd harmonics are emergence due to the slit gap $\Delta t$ is a half cycle and the dynamic phase difference $\Theta$ vanish between each two emission channels. One can observe that the model of the temporal two-slit interference reproduces the peaks of $q$-th harmonic and spectral splittings, shown by the white curve in Fig. \ref{Fig1}(a). Therefore, HHGs in the plateau zone will be divided into two zones which correspond to the interference opening and closing zones, as presented by the shadow and bright zones in Fig. (\ref{Fig3}) respectively.

Finally, the frequency redshift in the second feature will be explained qualitatively. The time-frequency analysis of the HHG with a longer dephasing time in the Fig. \ref{Fig1}(d) shows that more harmonics generate on the downhill part of the pulse than on the uphill part, which gives rise to the redshifts. The frequency redshift implies that the effect of dephasing in solids regulates a difference of the ionized dynamics between the uphill and downhill part of the pulse \cite{McDonald}.

\begin{figure}[htbp]
\centering
\includegraphics[width=8.5 cm,height=5 cm]{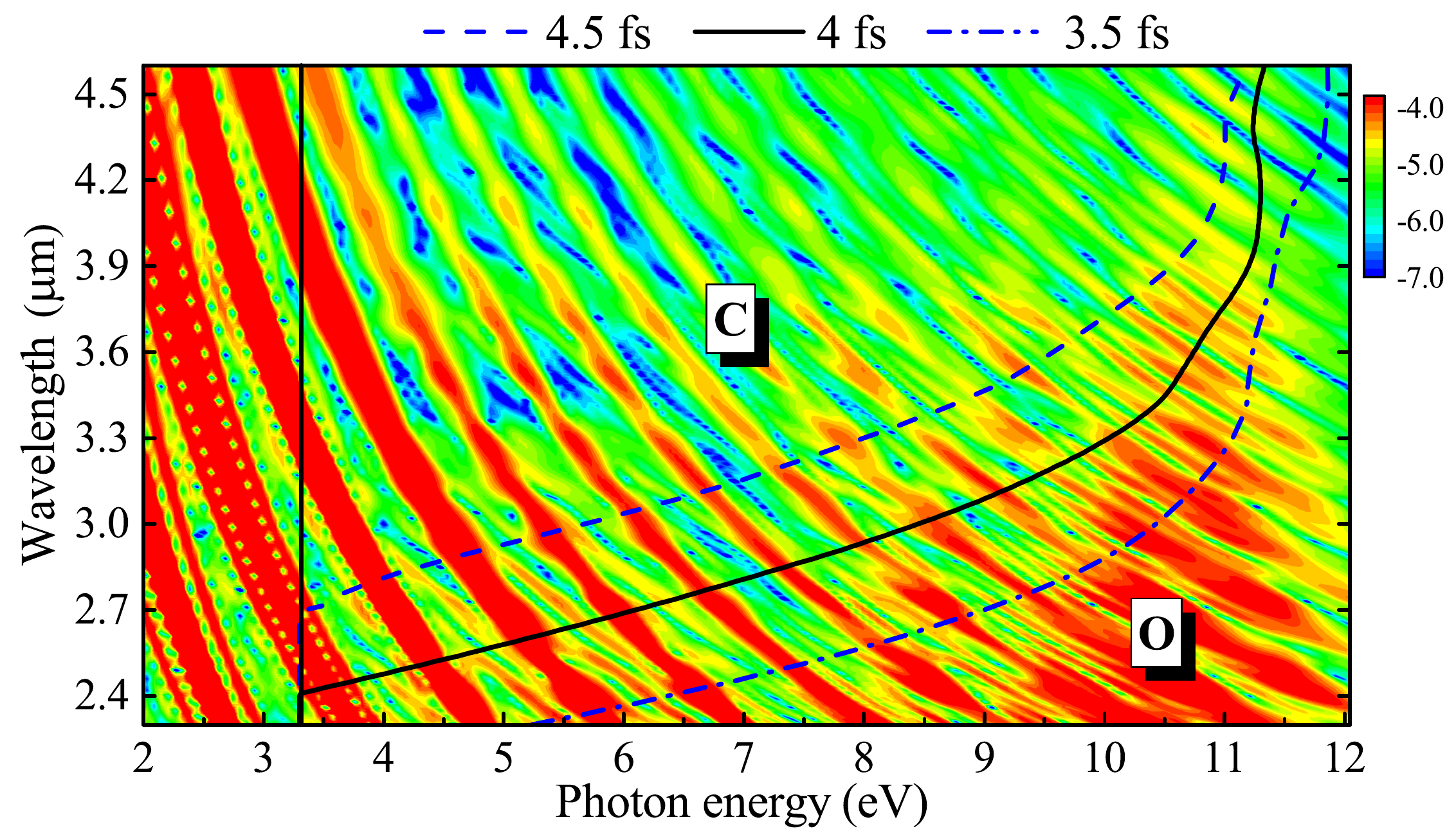}
\caption{(Color online) High-harmonic spectra with varying the driving wavelength. The laser vector is same as Fig. (\ref{Fig1}) and equals to 0.8$\pi$/a. The dephasing time $T_{2}$ = 4 fs is adopted. The vertical solid line shows the minimal band gap between conduction and valence bands. A black solid curve predicted by Eq. (\ref{E14}) shows the boundary with the used dephasing time 4 fs and distinguish the interference opening (\textbf{O}) and closing (\textbf{C}) zones. The blue dash and dash-dotted curves are the boundaries with two hypothetical dephasing times 4.5 and 3.5 fs respectively. The color scale is logarithmic.}\label{Fig4}
\end{figure}

From the above, the role of the dephasing time in the HHG from solids have been investigated. Within the density matrix formalism, phase relaxation accounting for electron-particle interactions are included by dephasing term and lead to the irreversible decay of quantum coherence and state population. Thus, measurement of characteristic time for the coherent processes is essential for both fundamental and applied research. The dephasing time $T_{2}$ in crystals at room temperature has been deemed that it is comparable to typical optical period of the MIR laser pulse \cite{Kruchinin,Sergaeva}. It implies that this characteristic time can be exposed by the MIR laser pulses. Based on the clarified mechanisms, the spectral feature of HHG would be used to retrieve the dephasing time $T_{2}$ of crystals accurately.

In Fig. (\ref{Fig4}), high-harmonic spectra with the varying driving wavelength have been obtained and can be separated into the interference opening (\textbf{O}) and closing (\textbf{C}) zones. At the plateau zone with photon energy ranging from 3.3 to 12 eV, one can observe that the harmonic peaks become more clearer with increasing wavelength. As a consequence, the emergence and absence of the spectral interference are distinguished delicately by a black solid curve in Fig. (\ref{Fig4}) and successfully predicted by the Eq. (\ref{E14}).  To make an assessment of this resultant boundary, two boundaries predicted via assuming the dephasing time with 4.5 and 3.5 fs are shown by the blue dash and dash-dotted curves respectively, which are obviously different from the black solid curve. One could conclude that the determination of boundary gives rise to an avenue to retrieve the dephasing time at subfemtosecond resolution. In addition, one may further find that the harmonic yield in the plateau zone are modulated with growing wavelength in Fig. (\ref{Fig4}), which can be attributed to the subcycle interference between two adjacent Zener events within an optical cycle and be discussed in Ref. \cite{Du5}. Mechanisms of the harmonic yield modulation had guided us toward the retrieve of the dynamic phase of Bloch electron \cite{Du5,Du6}. It is also worthwhile to note that this theoretical detection scheme with varying wavelength is not limited to MIR laser field, thereby being feasible to terahertz field and other laser parameters. 

\section{CONCLUSION}
To conclude, we have investigated the impact of dephasing time in the HHG from solid crystals by solving the density matrix equations. We found two features in the HHG spectra with growing dephasing time: (i) the yield of certain harmonic enhances monotonously while HHG spectra show an anomalously significant enhancement for the harmonics with higher order $q$; (ii) spectral splitting and frequency redshift. The coherent decay caused by the electron-particle scatterings in solids leads to the emergence of these two spectral features which further pave a way to retrieve the dephasing time at a subfemtosecond resolution by distinguishing the boundary line between the spectral interference opening and closing zones. The peaks of spectral splitting and $q$-th harmonic are predicted by the temporal two-slit interference model. In addition, we also provide a realizable scheme in present experimental setups to probe the dephasing time by measuring the wavelength-dependent solid HHG spectra. Measurement of the dephasing time can provide a new sight into microscopic interaction information in solids that the existing methods cannot reveal. It further advances the understanding of the ultrafast modulation of light, which has potential applications in petahertz electronic signal processing or strong-field optoelectronics.

\section*{ACKNOWLEDGMENTS}
The author thanks Hui-Hui Yang very much for helpful discussions. This work is supported by the National Natural Science Foundation of China (NSFC) (Grant No. 11904331).

\section*{APPENDIX A: Energy bands and \emph{k}-dependent dipole elements for \emph{w}-ZnO crystal}
\begin{figure}[htbp]
	\centering
	\includegraphics[width=8.5 cm,height=4 cm]{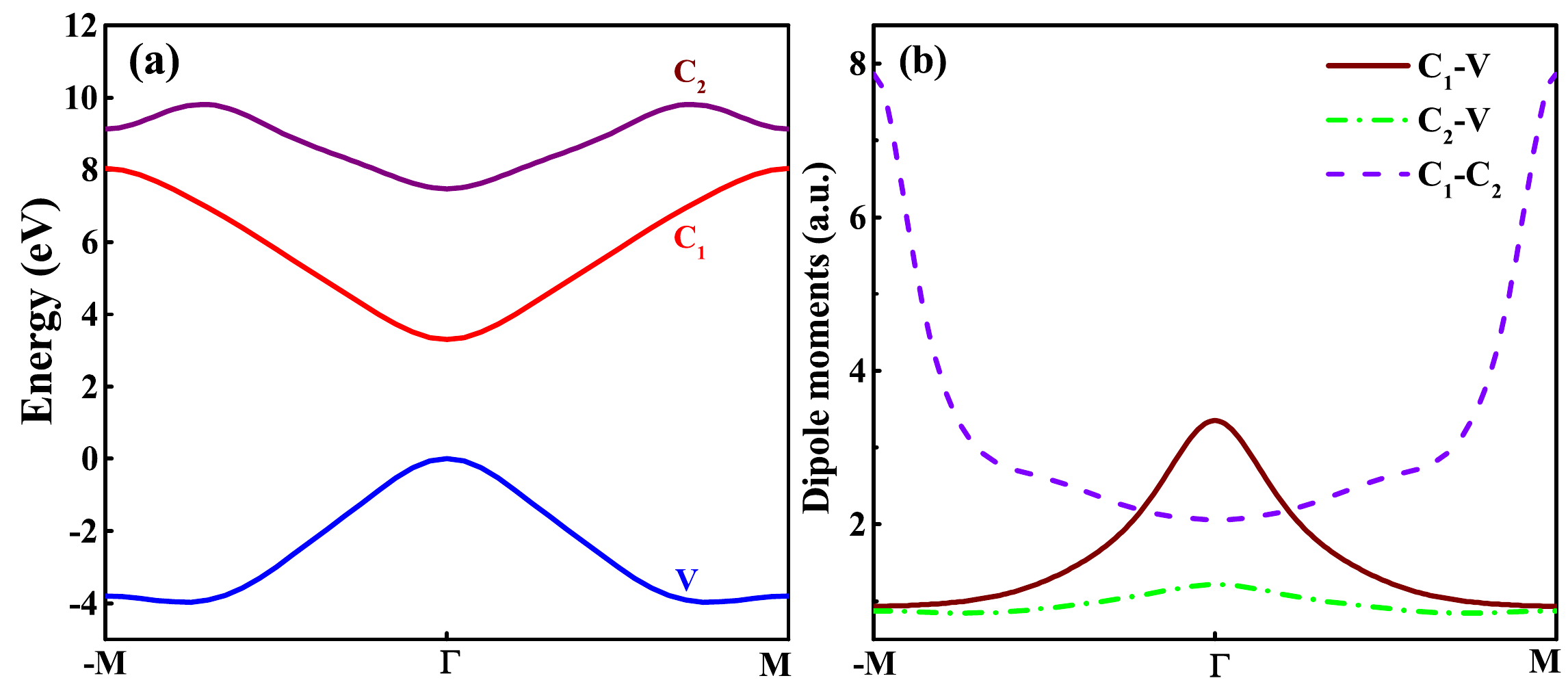}
	\caption{(Color online) (a) Energy band structure along the $\Gamma M$ direction.  (b) The transition dipole elements between each pair of bands considered.}\label{Fig5}
\end{figure}

 Accurate band structure and transition dipole moments are required before we perform the simulations of the light-solid interaction. These quantities are obtained via the density functional theory with a plane-wave basis set, using the Vienna ab initio simulation package (VASP) code \cite{Kresse,Perdew}. The equilibrium structure calculations of the \emph{w}-ZnO crystal are performed within the generalized gradient approximation (GGA) in the form of the Perdew and Wang (PW91) functional which is used to approximate exchange and correlation potentials. The cutoff energy of 550 eV is employed for the plane-wave basis expansion. The convergence criteria for the total energies and ionic forces are set to be $10^{-8}$ eV and $10^{-5}$ eV\AA$^{-1}$ in the formula unit. The Brillouin zone of the unit cell has been sampled by a 13$\times$13 $\times$8 Monkhorst-Pack  \emph{k}-point mesh for self-consistent converged calculations. By using the the Rayleigh-Schr\"{o}dinger perturbation theory, the \emph{k}-dependent transition dipole moments are related to the momentum elements $\hat{p}_{mm'}(k)$ as
\renewcommand\theequation{A1}
\begin{equation*}\label{A1}
	\begin{split}
		& \textbf{d}_{m'm}(k) = i\cdot\frac{\hat{p}_{m'm}}{E_{m'}(k)-E_{m}(k)}, \\
		& \hat{p}_{m'm}(k) = \langle u_{m',k}(r)|\hat{p}|u_{m,k}(r)\rangle,
	\end{split}
\end{equation*}
where $|u_{m,k}(r)\rangle$ is the periodic part of the Bloch function in band index $m$ with crystal momentum \emph{k}. Once the Bloch function is calculated, the value of the transition dipole elements can also be calculated. The calculated energy bands are fitted with the function $E_{m}(k) = \sum _{i=0}^{\infty} \alpha_{m,i} \cos(ika)$. The coefficients are listed in TABLE I.  $a = 5.32 $ a.u is the lattice constant along the $\Gamma M$ direction (optical axis). The energy bands and transition dipole elements $\textbf{d}(k)$ presented in Fig. \ref{Fig5} show a good agreement with the results in Refs. \cite{Vampa1,Du5,Vampa5}. 

\begin{table}
	\caption{Coefficients of the expansion of the energy bands.}
	\centering
	\begin{tabular}{p{1.5cm}<{\centering} p{2cm}<{\centering} p{2cm}<{\centering} p{2cm}<{\centering}}	
		\hline \hline
		ZnO   &V   &$C_{1}$   &$C_{2}$\\
		\hline
		$\alpha_{0}$  &-0.0928   &0.2111    &0.3249   \\
		$\alpha_{1}$   &0.0705  &-0.0814    &-0.0363   \\
		$\alpha_{2}$   &0.0200  &-0.0024    &-0.0146   \\
		$\alpha_{3}$   &-0.0012   &-0.0048  &0.0059   \\
		$\alpha_{4}$   &0.0029   &-0.0003   &-0.0050   \\
		$\alpha_{5}$   &0.0006   &-0.0009   & 0.0000   \\		
		\hline   \hline
	\end{tabular}		
\end{table}

\begin{figure}[htbp]
	\centering
	\includegraphics[width=8.5 cm,height=4 cm]{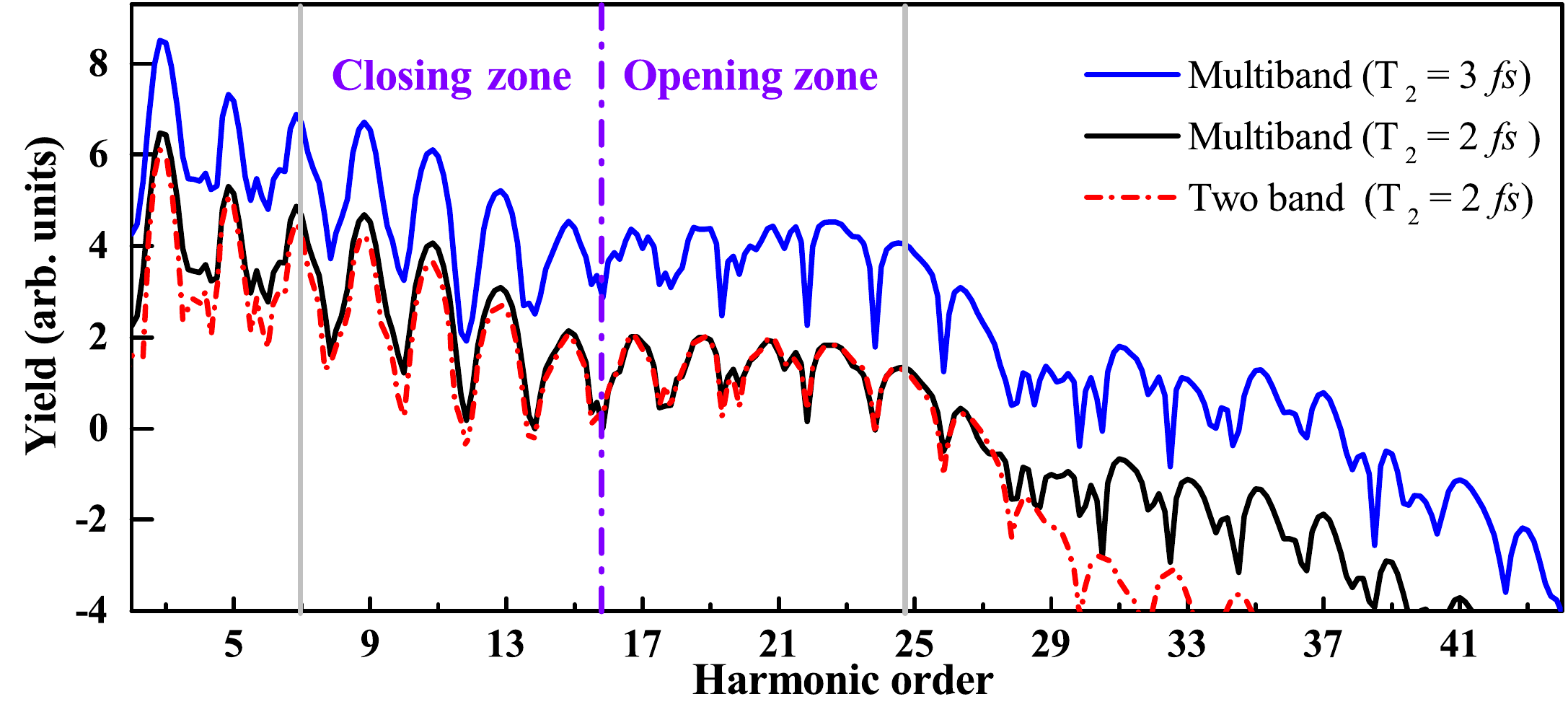}
	\caption{(Color online) Effect of the multiband and higher laser intensity. The laser vector adopted here is $\pi/a_{0}$ which corresponding to the higher laser intensity 4.81 TW cm$^{-2}$. }\label{Fig6}
\end{figure}

\section*{APPENDIX B: Multiband and higher-laser-intensity effects}
In this appendix, we make an assessment on the role of multiband and higher laser intensity in the HHG spectrum. To exclude the complex channel interference induced by the Bloch-Zener oscillation across the BZ edge, the vector of laser field adopted in this work is limited within $\pi/a_{0}$. One can find that the HHG spectra (time-frequency analyses) of the primary plateau simulated by the two band model shows a good agreement with the results of multiband model, as shown by the red dash-dotted and black solid curves respectively in Fig. (\ref{Fig6}). Time-frequency analyses are not shown here. In the multiband simulations with a higher laser intensity, the discussed spectral feature can be observed clearly. As shown by the blue solid curve compared with the black solid curve in Fig. (\ref{Fig6}), the HHG spectrum with a longer dephasing time (3 fs) shows two distinguished interference closing and opening zones.


\begin{thebibliography}{}	
\bibitem{Ferray} M. Ferray, A. L'Huillier, X. F. Li, L. A. Lompre, G. Mainfray, and C. Manus, J. Phys. B {\bf 21}, L31 (1988).
\bibitem{Hentschel} M. Hentschel, R. Kienberger, C. Spielmann, G. A. Reider, N. Milosevic, T. Brabec, P. Corkum, U. Heinzmann, M. Drescher, and F. Krausz, Nature {\bf 414}, 509 (2001).
\bibitem{Smirnov} O. Smirnova, Y. Mairesse, S. Patchkovskii, N. Dudovich, D. Villeneuve, P. Corkum, and M. Y. Ivanov, Nature {\bf 460}, 972 (2009).
\bibitem{Worner} H. J. W\"orner, J. B. Bertrand, D. V. Kartashov, P. B. Corkum, and D. M. Villeneuve, Nature {\bf 466}, 604 (2010).
\bibitem{Peng} L.-Y. Peng, W.-C. Jiang, J.-W. Geng, W.-H. Xiong, and Q. Gong, Phys. Rep. {\bf 575}, 1 (2015).	
\bibitem{Kruchinin} S. Y. Kruchinin, F. Krausz, and V. S. Yakovlev, Rev. Mod. Phys. {\bf 90}, 021002 (2018).
\bibitem{Pronin} K. A. Pronin and A. D. Bandrauk, Phys. Rev. Lett. {\bf 97}, 020602 (2006).
\bibitem{Ghimire} S. Ghimire, A. D. DiChiara, E. Sistrunk, P. Agostini, L. F. DiMauro, and D. A. Reis, Nat. Phys. {\bf 7}, 138 (2011).
\bibitem{Vampa} G. Vampa, T. J. Hammond, N. Thir\'e, B. E. Schmidt, F. L\'egar\'e, C. R. McDonald, T. Brabec, and P. B. Corkum, Nature {\bf 522}, 462 (2015).
\bibitem{Ndabashimiye} G. Ndabashimiye \emph{et al.}, Nature {\bf 534}, 520 (2016).
\bibitem{You_NC} Y. S. You \emph{et al.}, Nat. Commun. {\bf 8}, 724 (2017).
\bibitem{Liu_hanzhe} H. Liu Y. Li, Y. S. You, S. Ghimire, Tony F. Heinz, and D. A. Reis, Nat. Phys. {\bf 13}, 262 (2017).
\bibitem{Yoshikawa} N. Yoshikawa, T. Tamaya, and K. Tanaka, Science {\bf 356}, 736 (2017).
\bibitem{You_NP} Y. S. You, D. A. Reis, and S. Ghimire, Nat. Phys. {\bf 13}, 345 (2017).
\bibitem{Luu} T. T. Luu, M. Garg, S. Yu. Kruchinin, A. Moulet, M. Th. Hassan, and E. Goulielmakis, Nature {\bf 521}, 498 (2015).
\bibitem{Vampa_recon} G. Vampa, T. J. Hammond, N. Thir\'e, B. E. Schmidt, F. L\'egar\'e, C.R. McDonald, T. Brabec, D. D. Klug, and P. B. Corkum, Phys. Rev. Lett. {\bf 115}, 193603 (2015).
\bibitem{Banks_Berry} H. B. Banks, Q. Wu, D. C. Valovcin, S. Mack, A. C. Gossard, L. Pfeiffer, R.-B. Liu, and M. S. Sherwin, Phys. Rev. X {\bf 7}, 041042 (2017). 
\bibitem{Luu_Berry} T. T. Luu and H. J. W\"orner, Nat. Commun. {\bf 9}, 916 (2018).
\bibitem{Yangfang1} F. Yang and R.-B. Liu, Phys. Rev. B {\bf 90}, 245205 (2014).
\bibitem{Yangfang2} F. Yang, X. Xu and R.-B. Liu, New J. Phys. {\bf 16}, 043014 (2014).
\bibitem{Yangfang3} F. Yang and R.-B. Liu, New J. Phys. {\bf 15}, 115005 (2013). 
\bibitem{Du1} T.-Y. Du and X. B. Bian, Opt. Express {\bf 25}, 151 (2017).
\bibitem{Hawkins} P. G. Hawkins, M. Y. Ivanov, and V. S. Yakovlev, Phys. Rev. A {\bf 91}, 013405 (2015). 
\bibitem{Du3} T.-Y. Du, D. Tang, X. H. Huang, and X. B. Bian, Phys. Rev. A {\bf 97}, 043413 (2018).
\bibitem{Ikemachi} T. Ikemachi, Y. Shinohara, T. Sato, J. Yumoto, M. Kuwata-Gonokami, and K. L. Ishikawa, Phys. Rev. A {\bf 95}, 043416 (2017).
\bibitem{Du2} T.-Y. Du, X. H. Huang, and X. B. Bian, Phys. Rev. A {\bf 97}, 013403 (2018).
\bibitem{Rossi1} F. Rossi, Semicond. Sci. Tech. {\bf 13}, 147 (1998).
\bibitem{Rossi2} F. Rossi and T. Kuhn, Rev. Mod. Phys. {\bf 74}, 895 (2002).
\bibitem{Landau}  L. D. Landau, Sov. Phys. JETP {\bf  7}, 203 (1937).
\bibitem{Kaganov} M. I. Kaganov, I. M. Lifshitz, and L. V. Tanatarov, Sov. Phys. JETP {\bf 4}, 173 (1957).
\bibitem{Vampa1} G. Vampa, C. R. McDonald, G. Orlando, D. D. Klug, P. B. Corkum, and T. Brabec, Phys. Rev. Lett. {\bf 113}, 073901 (2014).
\bibitem{Du5} T.-Y. Du, D. Tang, and X. B. Bian, Phys. Rev. A {\bf 98}, 063416 (2018).
\bibitem{Vampa5}  G. Vampa, C. R. McDonald, G. Orlando, P. B. Corkum, and T. Brabec, Phys. Rev. B {\bf 91}, 064302 (2015).
\bibitem{Vampa4} G. Vampa and T. Brabec, J. Phys. B {\bf 50}, 083001 (2017).

\bibitem{zhangxiao} X. Zhang, J. Li, Z. Zhou, S. Yue, H. Du, L. Fu, and H.-G. Luo, Phys. Rev. B {\bf  99}, 014304 (2019).
\bibitem{Huangkun} K. Huang, \textit{Solid State Physics} (Higher Education Press, Beijing, China, 1988).
\bibitem{Ashcroft} N. W. Ashcroft and N. D. Mermin, \textit{Solid State Physics} (Philadelphia, PA: Saunders, 1976).
\bibitem{Du_inhomo} T.-Y. Du, Z. Guan, X.-X. Zhou, and X.-B. Bian, Phys. Rev. A  {\bf 94} 023419 (2016).
\bibitem{Lewenstein} M. Lewenstein, P. Sali\'eres, and A. L'Huillier, Phys. Rev. A {\bf  52}, 4747 (1995). 
\bibitem{Chandre} C. Chandre, S. Wiggins, and T. Uzer, Physica D {\bf 181}, 171 (2003).
\bibitem{McDonald} C. R. McDonald \emph{et al.},  J. Opt. {\bf 19} 114005 (2017).
\bibitem{Sergaeva} O. Sergaeva, V. Gruzdev, D. Austin, and E. Chowdhury, J. Opt. Soc. Am. B {\bf 35}, 2895 (2018).

\bibitem{Du6} T.-Y. Du and S.-J. Ding, Phys. Rev. A {\bf 99}, 033406 (2019).
\bibitem{Kresse} G. Kresse and J. Furthm\"uller, Phys. Rev. B {\bf 54}, 11169 (1996).
\bibitem{Perdew} J. P. Perdew, K. Burke, and M. Ernzerhof, Phys. Rev. Lett. {\bf 77}, 3865 (1996).


\end{thebibliography}
\end{document}